\documentclass[12pt]{article}
\usepackage{amssymb,amsmath}
\begin{document}


\sloppy
\title
{\Large On the signature of pressure in  gravity   }

\author
 {
       A.I.Nikishov
          \thanks
             {E-mail: nikishov@lpi.ru}
  \\
               {\small \phantom{uuu}}
  \\
           {\it {\small} I.E.Tamm Department of Theoretical Physics,}
  \\
               {\it {\small} P.N.Lebedev Physical Institute, Moscow, Russia}
  \\
 }
%
\maketitle



\begin{abstract}
When pressure is not negligible in comparison with energy density, the external gravitational field and the motion of particles in it are modified. For spherically 
symmetric body two effective mass parameters determine the external gravitational 
field and the motion of particles in it. For distances much larger then the gravitational radius we use the linearized Einstein equations to consider the effects of pressure on test particle  motion.
\end{abstract}

\section{Introduction}
For strongly compressed star the pressure may noticeably influence the gravitational 
field not only inside the star but also outside if it [1] In the region $\frac{r_g}{r}<<1$ it is sufficient to use the linearized Einstein equation to consider 
the influence of pressure on gravitational field. This is done in Section 2. In Section 3 we  approach the problem using Schwinger's  Theory of Sources to make sure 
that it gives the same results as we obtain in this paper. In section 4 we consider the influence of pressure on the motion of test particle.

\section{Pressure in gravity. } 
We assume that the energy-momentum tesor of the gravitating body is

$$
T^{\mu\nu}=\rm diag(\epsilon, p,p,p),                        \eqno(1)
$$
The linearized Einstein equation may be written in the form $(g_{\mu\nu}=\eta_{\mu\nu}+h_{\mu\nu})$
$$
\Delta h_{\mu\nu}=-\frac{16\pi G}{c^4}\bar T_
{\mu\nu},\quad \bar T_{\mu\nu}=T_{\mu\nu}-\frac12\eta_{\mu\nu}T_l{}^l, \quad
\eta_{\mu\nu}=\rm diag(-1,1,1,1.),                                         \eqno(2)
$$
Here
$$
\bar T_{00}=\frac12(\epsilon+3p),\quad
\bar T_{11}=\bar T_{22}=\bar T_{33}=\frac12(\epsilon-p).                       \eqno(3)
$$

The solution of (2) is well known, 
$$
h_{\mu\nu}(\vec x)=4\frac{G}{c^4}\int\frac{\bar T_{\mu\nu}(\vec x')}{|\vec x-\vec x'|}d^3x'.                                                                                \eqno(4)
$$
 For spherically symmetric body it follows from (4) that outside the body
$$
h_{00}=\frac{2Gm_t}{rc^2}, \quad h_{s}\equiv h_{11}=\frac{2Gm_s}{rc^2}. \eqno(5)
$$
Here $m_t$ is the effective mass (with the pressure taken into account) defining $h_{00}$ and
similarly for space part of metric $h_s\equiv h_{11}$:
$$
m_t=c^{-2}\int[\epsilon+3p]d^3x, \quad m_s=c^{-2}\int[\epsilon-p]d^3x. \eqno(6)
$$
The subscripts $t$ and $s$ remind us about time- and space parts of the metric.
We also define
$$
r_{gt}=\frac{2Gm_{t}}{c^2}, \quad r_{gs}=\frac{2Gm_{s}}{c^2}. \eqno(7)
$$
In spherical coordinate system we get 

$$
-ds^2=g_{00}(dx^0)^2 + g_{s}(dr^2+r^2d\theta^2+r^2\sin^2\theta d\varphi^2),
$$
$$
g_{00}=-(1-h_{00})=-(1+\frac{2\varphi}{c^2}),\quad g_s\equiv g_{11}=1+h_s=1+\frac{2Gm_s}{rc^2},                                                                               \eqno(8) 
$$
see (5).
The equation of motion for radial coordinate is
$$
\frac{d^2r}{ds^2}=-\Gamma^r_{\mu\nu}\frac{dx^{\mu}}{ds}\frac{dx^{\nu}}{ds}, \quad x^0=ct.  \eqno(9)
$$
For particle at rest 
$$
\frac{d^2r}{ds^2}=-\Gamma^r_{00}=\frac12g^{rr}\frac{\partial g_{00}}{\partial r}=
\frac12\frac{\partial h_{00}}{\partial r}.
                                                                          \eqno(10)
$$
With the help of (5) we find
$$
\frac{d^2r}{dt^2}=-\frac{\partial \varphi}{\partial r}, \quad \varphi=-\frac{Gm_t}{r}.
                                                                           \eqno(11)
$$
In contrast to Schwarzschid solution the parameter $m_t$ depends on pressure. In view
of this it is instructive to make a short excursion to Schwiger's Theory of Sources [2].  
\section{Pressure in Schwinger's  Theory of Sources}

We denote Schwinger's total energy-momentum tensor $T^{\mu\nu}$ in eq. (4.34) Ch.2 in
[2] as $\theta^{\mu\nu}$and write it in the form
$$
E(x^0)=-\frac{G}{c^4}\int\frac{d^3xd^3x'}{|\vec x-\vec x'|}[\theta^{\mu\nu}(\vec x,x^0)\theta_{\mu\nu}(\vec x',x^0)-\frac12\theta(\vec x,x^0)\theta(\vec x',x^0)],\quad \theta=\theta^{\nu}{}_{\nu}.                                 \eqno(12)
$$
Using $\theta^{\mu\nu}=T^{\mu\nu}+t^{\mu\nu}$ we get from (12) the interaction energy
$$
E_{int}(x^0)=-\frac{G}{c^4}\int\frac{d^3xd^3x'}{|\vec x-\vec x'|}\{[T^{\mu\nu}(\vec x,x^0)t_{\mu\nu}(\vec x',x^0)+(\vec x\leftrightarrow\vec x')]
$$
$$-\frac12[T(\vec x,x^0)
t(\vec x',x^0)+(\vec x\leftrightarrow\vec x')]\}.                         \eqno(13)
$$
Here $(\vec x\leftrightarrow\vec x')$ means terms obtained from the preceding ones by substitution$(\vec x\leftrightarrow\vec x')$.

I. We first assume that
$$
t^{\mu\nu}(\vec x.x^0)=m'c^2\delta_{\mu0}\delta_{\nu0}\delta(\vec x-\vec R),                                                                                      \eqno(14)
$$ 
i.e. the test particle is at rest. Then , from (13) and (14) we find
$$
E_{int}=-\frac{Gm'}{c^2}\int\frac{d^3x}{|\vec x-\vec R|}\{T^{00}(\vec x)-\frac12
[T^{00}-T_{kk}]+ (\vec x\leftrightarrow\vec x')\}=
$$
$$
-\frac{Gm'}{c^2}\int\frac{d^3x}{|\vec x-\vec R|}\frac12\{[\epsilon(\vec x)+3p(\vec x)]+ (\vec x\leftrightarrow\vec x')\}=-\frac{Gm'}{c^2}\int\frac{d^3x}{|\vec x-\vec R|}\{[\epsilon(\vec x)+3p(\vec x)].                                    \eqno(15)
$$

For spherically symmetric gravitating body we get
the Newtonian law with $m\to m_t$:
$$
E_{int}=-\frac{Gm'm_t}{R}=m'\varphi \eqno(16)
$$
in agreement with (11).

II. Next, we consider the interaction of a photon beam:
$$
t^{\mu\nu}=\sigma p^{\mu}p^{\nu},\quad p^2=-(p^0)^2+\vec p{}^2=0,
                                                                          \eqno(17)
$$
see eq. (4.38) in Ch.2 in [2].
Using (17) and (13) we obtain
$$
E_{int}=-\frac{G}{c^4}\int\frac{d^3xd^3x'}{|\vec x-\vec x'|}\sigma p^0\{[\epsilon(\vec x)+p(\vec x)]+(\vec x\leftrightarrow\vec x')\}=
-\frac{G}{c^4}\int\frac{d^3xd^3x'}{|\vec x-\vec x'|}\sigma2p^0\{[\epsilon(\vec x)+p(\vec x)],
                                                      \eqno(18)
$$
i.e. the interaction energy of the photon with the gravitating body is twice the Newtonian value (which is obtained by substituting $m'c^2\to p^0, m\to m_{eff}$),
cf. [2];
$$
m_{eff}=\frac{m_t+m_s}{2}.                   \eqno(19)
$$
It follows from here that the deflexion angle for the photon flying through the gravitational field of a spherically symmetric body is
$$
\theta=\frac{4Gm_{eff}}{\rho c^2}=\frac{2r_{eff}}{\rho},     \eqno(20)
$$
see eq. (4.41) in Ch2 in [2]. Here $\rho$ is the impact parameter. The important thing for us is that $m_{eff}\neq m_t$.
\section{Test particle in metric (8)}
We use the Hamilton-Jacobi equation, see \S 101 in [3]
$$
g^{\mu\nu}\frac{\partial S}{\partial x^{\mu}}\frac{\partial S}{\partial x^{\nu}}=
-m'^2c^2.                                                         \eqno(21)
$$
 With
$$
g^{00}=-(1-\frac{r_{gt}}{r})^{-1},\quad g_s{}^{-1}=(1+\frac{r_{gt}}{r})^{-1}
$$
we have
$$
(1-\frac{r_{gt}}{r})^{-1}(\partial S/\partial t)^2-(1+\frac{r_{gs}}{r})^{-1}
[(\partial S/\partial r)^2+r^{-2}(\partial S/\partial \theta)^2]=
m'{}^2c^2.                                                      \eqno(22)
$$
$S$ has the form
$$
S=-{\cal E}_0t+M\theta+S_r,                                    \eqno(23)
$$
where $M$ is the angular momentum.
Using this in (22), we find
$$
S_r=\int\sqrt {F(r)}dr,\quad F(r)=\frac{{\cal E}_0^2}{c^2}\frac{1+\frac{r_{gs}}{r}}
{1-\frac{r_{gt}}{r}}-\frac{M^2}{r^2}-m'{}^2c^2(1+\frac{r_{gs}}{r}).       \eqno(24)
$$
The function $r=r(t)$ is given by the condition $\frac{\partial S}{\partial {\cal E}_0}=const$ and the trajectory is obtained from $\frac{\partial S}{\partial M}=const$,
see \S 101 in [3]:
$$
t=\frac{{\cal E}_0}{c^2}\int \frac{1+\frac{r_{gs}}{r}}
{1-\frac{r_{gt}}{r}} F^{-\frac12}dr,                            \eqno(25)
$$
$$
\theta=\int\frac{M}{r^2}F^{-\frac12}dr,                                   \eqno(26)
$$
In the nonrelativistic limit $c\to\infty$ we have
$$
  \frac{{\cal E}_0^2-m'{}^2c^4}{c^2}\approx E_{nonrel}2m',.         
$$
$$
F(r)\approx\frac{{\cal E}_0^2}{c^2}(1+\frac{r_{gs}+r_{gt}}{r})
-\frac{M^2}{r^2}-m'{}^2c^2(1+\frac{r_{gs}}{r})\approx                    \eqno(27)     
$$
$$
2m'(E_{nonrel}-E_{int})-\frac{M^2}{r^2},\quad E_{int}=-\frac{Gm'm_t}{r}. \eqno(28)
$$
Using (27), (28) in eqs. (25) and (26) we get the equations for the nonrelativistic particle. This agrees with (11) and (16).

For $m'\to0$  we see from (25), (26) that in the first approximation the role of
$m$ plays $m_{eff}=(m_s+m_t)/2$ in agreement with (19) 

Starting from (26) with $m'=0$ it is not difficult to obtain corrections to the leading
term (20) in the form of powers of $1/\rho$.

In terms of
$$
\delta=\frac{r_{gt}}{\rho},\quad r_{gs}=\zeta r_{gt},    \eqno(29)
$$
where $\rho$ is the impact parameter, we have instead (26)
$$
\theta=\int\sqrt{\frac{1-u\delta}{f(u)}}du,\quad f(u)=1-u^2+(u\zeta+u^3)\delta =(u-u_1)(u-u_2)(u-u_3)\delta.    \eqno(30)
$$
For photon $M=\rho{\cal E}_0/c.$ and $\theta$ at half of the trajectory is, cf. [4]
$$
\theta_{\frac12}=(u_3\delta)^{-\frac12}\int_0^{u_2}\sqrt{\frac{1-u\delta}{1-uu_3^{-1}}}
R(u)^{-1/2}du,\quad R(u)=(u_2-u)(u-u_1).                             \eqno(31)
$$ 
This equation coinsides with eq. (42) in [4] and $f(u)$ is the same as that in eq.(2)
in [4]. So, using the results obtained there, it is easy to get the deflexion      angle for photon
$$
\theta=2\bigl(\theta_{\frac12}-\frac{\pi}{2}\bigr)=(1+\zeta)\delta+\pi\frac{1+\zeta}{2}\delta^2+\cdots.                     \eqno(32)
$$ 
The leading term
$$
(1+\zeta)\delta=\frac{r_{gt}+r_{gs}}{\rho}=\frac{4Gm_{eff}}{\rho c^2}
$$
agrees with (20). Here eqs. (7), (19) and (29) were used.

\section{Conclusion}
When pressure becomes noticeable, two mass parameters $m_t$ and $m_s$ differ from one another. In this case the Birkhoff theorem can be valid only with accuracy of order 
$\frac{m_t-m_s}{m_t}$ because at the beginning of contraction $m_t=m_s=m$.  In the considered approximation the presence of pressure increases the pull towards the gravitating body more for nonrelativistic particle
then for ultrarelativistic one.
 \section*{Acknowledgements}

The work was supported by Scientific Schools and Russian Fund for Fundamental Research (Grants 1615.2008.2 and 08-02-01118).

 \section*{References}
1. A.I.Nikishov,[gr-qc] 0912.5180\\
2. Schwinger J., {\sl Particles, Sources, and Fields}. Addison-Wesley, 1970. V. 1. \\
3. L.D.Landau and E.M.Lifshitz, {\sl The classical theory of
   fields}, Addison-Wesley, Cambridge, MA, 1971).\\
4. A.I.Nikishov, Part. Nucl, 2009, Vol.6, N 6, p704-716.
\end{document}